# Fast polymerization at low temperature of an infrared radiation cured epoxy-amine adhesive

Sébastien GENTY [a, b], Philippe TINGAUT [b], Maëlenn AUFRAY [a, *]

[a] CIRIMAT, Université de Toulouse, CNRS, Toulouse Cedex 4, France
[b] SOCOMORE, ZI du Prat, 56037 Vannes Cedex, France
*Corresponding author: maelenn.aufray@ensiacet.fr

## Abstract

In the industry, the cure time of two-component adhesives is very important for a cost-effective manufacturing. Too fast, it does not favor the application of the product and the control of bonded joints. Too slow, it leads to long process times and too high process costs. The best compromises are two-component adhesives that cure slowly at room temperature and can reach full polymerization in minutes, on demand. In this paper, the curing behavior of a model poly-epoxide adhesive (a stoichiometric mixture of a pure epoxy and amine) polymerized with infrared radiation will be studied. The kinetic follow-up of this polymerization will be carried out by thermal analysis (determination of the residual heat peak by Differential Scanning Calorimetry-DSC). This study paves the way to a cold and universal cure-on-demand process, which means achieved in few minutes at low temperature without any initiators, catalysts or accelerators. Basically, infrared curing can be possible thanks to an increase in temperature (called thermal effect). But it has been shown that a "non-thermal effect" could also be involved in accelerating kinetics with infrared. This increase due to a non-thermal effect, suggested as a function of the infrared radiative flux, has been shown to be possible thanks to the absorption of infrared radiation, leading to a reduction in the energy barrier of the primary epoxy/amine reaction.

## 1. INTRODUCTION

Cost and weight reductions are the two-main drivers in the aerospace industry, leading to the increase in the use of composite materials and structures. As an example, the Airbus A 350 and Boeing B 787 are composed of more than 50 % of composite materials (Carbon Fiber Reinforced Plastic - CFRP). From an industrial standpoint, the structure assembly is technically challenging. During the manufacturing process of composite materials, a permanent deformation (spring back) and thickness variation can be induced [1,2], so that a gap comes forth in the joining interface. In order to fill gaps and reduce stress concentration around rivets and fasteners, liquid shims are generally used. They are epoxy-based and bi-component adhesives which cure at room temperature and can be characterized by various properties (density, rheological and cure profile, adherence). Nowadays, research needs and improvements concern mainly a curing profile suitable to the manufacturing requirements and specificities. Thus, liquid shims should cure slowly at room temperature for purposes of extend the application time (also called handling time) and should be fast cured in few minutes once cast in place. This is only possible if a Shim Cured-On-Demand (SCOD), activated by an external way, is considered. The cure -on-demand of liquid shim can be under Ultra-Violet [3–8], microwaves [9–13] or magnetically induced [6,14–16], but only if additives are added to the liquid shim formulation (like photo initiator or paramagnetic particles). Instead of these common ways of curing, this present study paves the way to a more universal way of curing, avoiding any additives: infrared-curing. This polymerization method has already been studied in the literature [17–23], but only when the temperatures induced by infrared radiation are very important. To our knowledge, low temperature curing by infrared radiation (*i.e.* about 50°C) has never been reported from a fundamental standpoint. On another note, the infrared lamp herein operated has already been commercially available for several years [24] and is well known by men of the art as a means of accelerating the curing of composite repair materials or sealants [25]. The effectiveness of such a polymerization is no longer to be proved.

## 2. EXPERIMENTAL

### 2.1. Substrate and adhesives

The substrate used is a multi-material composed of Aluminum Alloy 2024-T3 with a thickness of $1.000 \pm 0.005$ mm from Rocholl GmbH (used for its conductivity), covered with a Polyethylene (PE) film in order to avoid any interphase formation between the adhesive and the substrate [26–28].

The epoxy pre-polymer is a bisphenol A diglycidyl ether (BADGE, Epoxy Equivalent Weight = 171-175 g.eq⁻¹, D.E.R. ™ 332 from Dow Chemicals) and the hardener is the triethylenetetraamine (TETA, Amine Hydrogen Equivalent Weight = 24 g.eq⁻¹, D.E.H. ™ 24 from Dow Chemicals). The model adhesive used in this study was a stoichiometric mixture of BADGE with 13.9 parts per hundred grams of resin (*phr*) of TETA. For these proportions, the infinite glass transition temperature (*i.e.* when the conversion degree is 1.0) is 138 °C.

### 2.2. Differential Scanning Calorimetry (DSC)

The curing degree (here named as conversion) was followed by differential scanning calorimetry with a Mettler Toledo thermal analyzer, model DSC 1 (Viroflay, France) and was performed with the STARe® evaluation software. Indium was used to calibrate the temperature and enthalpy measurements. Sealed aluminum pans of 15-20 mg of mixture, under a constant flow of nitrogen were used with two approaches: an isothermal scanning method, applied for thermal-cured adhesive conversion determination (*in situ* approach) and a non-isothermal method (dynamic approach). First, the non-isothermal scanning method consists in the determination of the residual enthalpy of the reaction ($\Delta H_R$) normalized by the total enthalpy of the reaction ($\Delta H_T$) so that the conversion can be calculated from the ratio of these two terms (Equation 1).

$$X(t) = 1 - \frac{\Delta H_R}{\Delta H_T}$$
Equation 1

Both enthalpies of reaction, $\Delta H_R$ and $\Delta H_T$, are determined from -20 °C to 250 °C at 10 °C.min$^{-1}$. Then, assuming that the measured heat flow ($\Phi$) by DSC is proportional to the rate of kinetic process (dx/dt) [29], the non-isothermal scanning method consists simply in the heat flow integration (Equation 2).

$$X(t) = \frac{1}{\Delta H_T} \int_0^t \Phi \, . \, dt$$

<div align="right">Equation 2</div>

The *in-situ* approach (Equation 1) was used for measurements of thermal-cured systems conversions (in an oven) when the dynamic approach (Equation 2) was mainly used to calculate the conversions of systems cured under infrared lamp. Both methods have been verified to give the same conversions (taking into account the uncertainty of the measurements).

## 2.3. Infrared curing device

The fast curing technique was carried out using an infrared lamp supplied by SUNAERO (Lyon, France). It is already used industrially, in aeronautics, for the accelerated polymerization of sealant or for composites repair [24]. This lamp is initially supplied to increase the kinetics of polymerization only by radiation energy transfer. It consists of a transportable control panel and an infrared lamp emitter. The Infrared lamp is held horizontally at 20 cm by means of a tripod. The induced temperature, heat-up ramp and heat time are set up in the control panel so that the emitted flux is automatically regulated by the temperature measurement of the cured sample with a K thermocouple. In this study, the same curing profile has always been applied to the sample: from room temperature to 50 °C at 5 °C.min⁻¹, followed by a constant temperature of 50 °C throughout the experiment. The infrared spectrum emitted by the lamp has been measured by using a Bruker 80V Infrared spectrometer (Marne la Vallée, France) and carried out in the medium range (*i.e.* from 400 to 4,000 cm$^{-1}$) with a KBr separator and a DLaTGS (Deuterated Lanthanum $\alpha$-alanine-doped TriGlycine Sulfate) detector, and near infrared (*i.e.* from 4,000 to 15,000 cm$^{-1}$) with the same detector and a CaF$_2$ separator. The spectrum has been recorded with a resolution of 4 cm$^{-1}$, with only 1 scan to allow a higher temporal resolution (Figure 1).

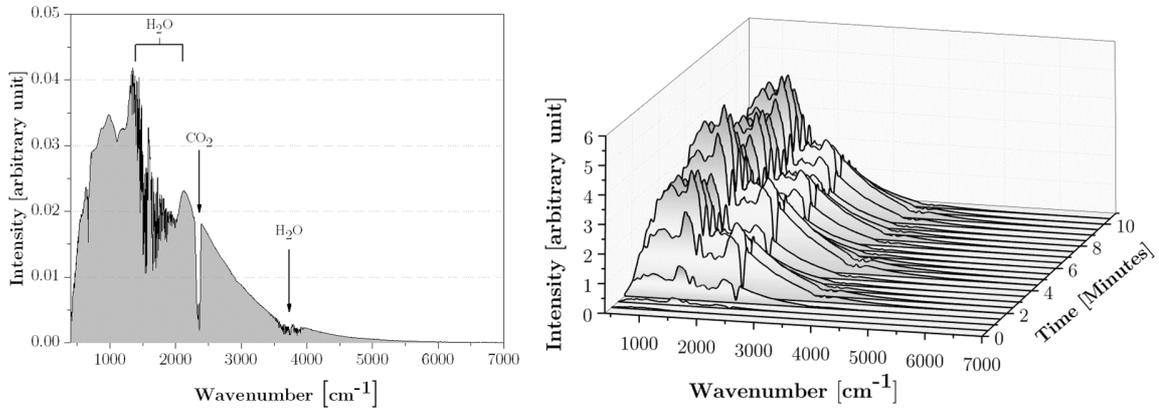

Figure 1: Infrared lamp emitter spectrum after 2 minutes (left) and in function of time (right).

The infrared emittance is mainly comprised in the medium range. The spectrum after 6,000 cm$^{-1}$ has not been plotted as it is close to zero. Some characteristic peaks are present at 2,367 cm$^{-1}$, (asymmetric elongation vibration of $CO_2$), between 1,400 and 2,000 cm$^{-1}$ and around 3,700 cm$^{-1}$ (respectively, the symmetrical/antisymmetric vibrations and the angle deformation of water). Thus, the infrared spectrum here presented is not the absolute emitted spectrum but rather as it is seen by the model adhesive.

The intensity of the infrared spectrum increases during the first 6 minutes so that the temperature reaches the set point (*i.e.* 50 °C) and then periodically varies around an average value to keep the induced temperature constant.

The infrared polymerization of the BADGE-TETA mixture was carried out for an induced temperature of 50 °C. As a reminder, the infrared energy emitted by the lamp is automatically regulated in order to reach and maintain the sample temperature as set up. Therefore, monitoring separately the induced temperature and the emitted radiative flux appears to be challenging. To address this issue, a specific test configuration has been elaborated. It consists in placing the substrate with the BADGE-TETA mixture onto a block of dry ice during the infrared curing. In other words, the IR lamp has been forced to emit a significant amount of energy in order to keep the adhesive at 50 °C. This polymerization has been named as IR-$\varphi_{Max}$. On the other hand, the original test configuration is an IR curing with an automatically-controlled radiative flux by the control panel, to maintain the temperature at 50 °C (named IR-$\varphi_{Auto}$). The radiative flux for these two curing configurations (IR-$\varphi_{Auto}$ and IR-$\varphi_{Max}$) and the temperature of the BADGE-TETA mixture were measured (Figure 2).

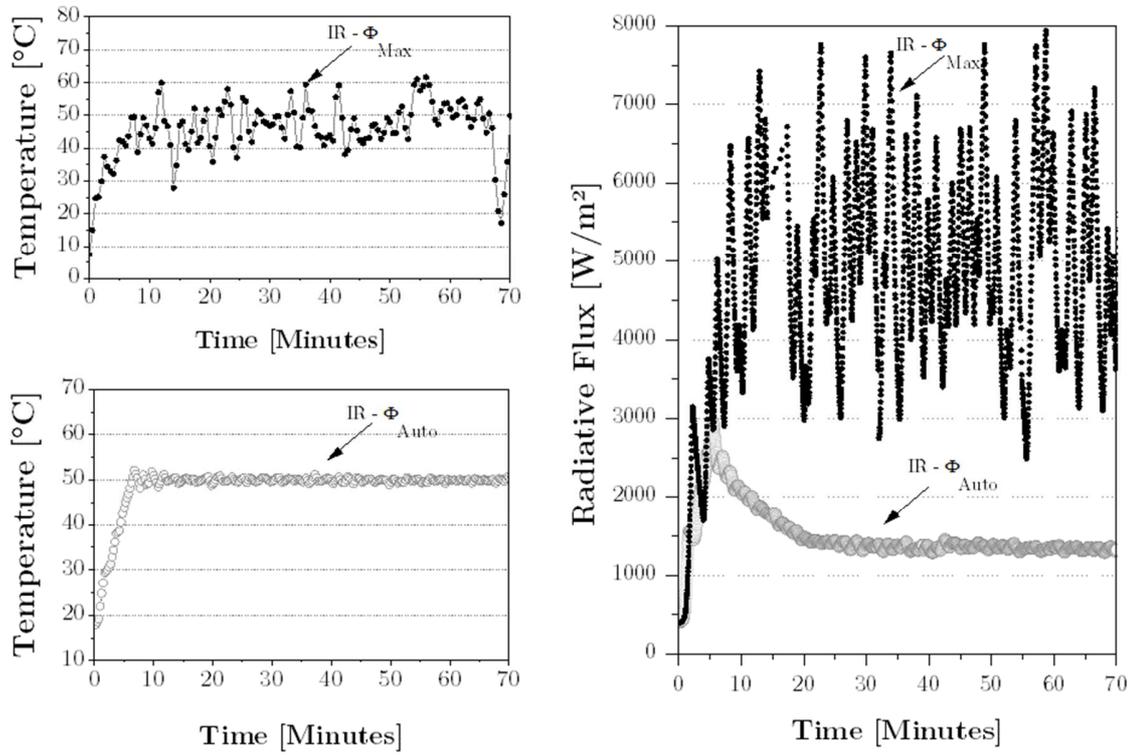

Figure 2: Temperature of the mixture (left) and radiative flux of the Infrared lamp emitter (right).

In the case of polymerization with the automatic adjusted radiative flux (IR-$\varphi_{Auto}$), the flux increases during the first 5 minutes until it reaches a maximum at 2,000 W/m² and then decreases over time to a constant value of 1,500 W/m² after 30 minutes of exposure. From there, the temperature and flow are constant. By considering the polymerization on dry ice (IR-$\varphi_{Max}$), the radiative flux increases in a non-continuous manner for the first 15 minutes and then varies periodically around an average value (approximately 5,000 W/m²) over the time, without any noticeable decrease in the trend.

# 3. RESULTS AND DISCUSSION

## 3.1. Influence of the radiative flux

The degree of cure of BADGE-TETA mixture under the two curing conditions (IR-$\varphi_{Auto}$ and IR-$\varphi_{Max}$), compared to the degree of curing at 50 °C (in an oven) is given in Figure 3.

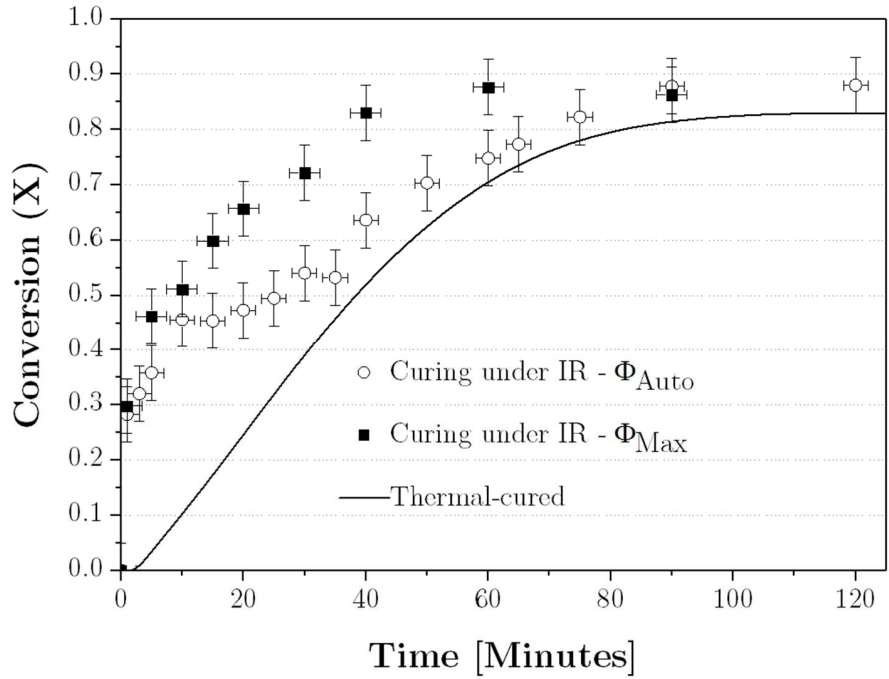

Figure 3: Conversions of BADGE-TETA mixture.

The time to gelation and vitrification obtained from the Figure 3 are summarized in Table 1. First, conversion at gel point can be theoretically determined according to the well-known theory of Flory-Stockmayer [30,31] reminded in the Equation 3, where $r$ is the stoichiometric ratio, $f$ and $g$ are the respective functionalities of the epoxide prepolymer and hardener.

$$\alpha_{gel} = \left[\frac{r}{(f-1) \times (g-1)}\right]^{\frac{1}{2}}$$

Equation 3

For BADGE-TETA mixture, $f = 2$ and $g = 6$ so that the prediction for the gel point conversion ($\alpha_{gel}$) is 0.45 for a stoichiometric mixture. Let us remind that it is commonly accepted that gelation has no effect on the curing rate (*i.e.* the conversion rate remains unchanged) [32]. Simply put, gelation time can be defined as the required time to reach out to a conversion of 0.45. Then, the glass transition temperature can be attributed to a conversion degree by using the Di Benedetto equation modified by Pascault [27] according to the Equation 4.

$$T_g(x) = T_{g0} + \frac{(T_{g\infty} - T_{g0})\lambda x}{1 - (1-\lambda)x}$$

Equation 4

The glass transition temperature of the system at the initial state ($t=0$) is $T_{g0}$ and the glass transition temperature of the fully cured system is $T_{g\infty}$. Pascault [33] named $\lambda$, the ratio between the variation in heat capacity at $T_{g0}$ and $T_{g\infty}$ according to Equation 5.

$$\lambda = \frac{\Delta C_{p\infty}}{\Delta C_{p0}} = \frac{T_{g\infty}}{T_{g0}}$$

Equation 5

According to the Equation 4, the conversion at vitrification is 0,70 ± 0,05 so that the time to vitrification is the time to get this conversion degree.

Table 1: Time to gelation and vitrification for thermal-cured, and IR-cured BADGE-TETA mixture.

| CURING | $t_{GEL}$ | $t_{GLASS}$ |
|---|---|---|
| **THERMAL** | 35 minutes | 60 minutes |
| **IR-$\varphi_{AUTO}$** | 10-20 minutes | 50 minutes |
| **IR-$\varphi_{MAX}$** | 5 minutes | 25 minutes |

While the temperature of curing is 50 °C for the three trials with infrared and either $\varphi_{Auto}$ or $\varphi_{Max}$, the cure kinetic is faster than pure thermal polymerization. Consequently, the curing speed under infrared is not only accelerated by the increase in induced sample temperature. An additional effect, which has been called "non-thermal effect" (as opposed to the thermal effect of infrared), was also brought to light.

Now, let us consider the infrared curing with $\varphi_{Auto}$. The accelerated cure profile slows down after 10 minutes, overlapping with the drop of the radiative flux (see the Figure 2). The potential relation between non-thermal effect and radiative flux is strengthened by the kinetics under IR-$\varphi_{Max}$, which exhibits a greater "non-thermal" effect than that without the radiative flux optimization. On the other hand, the kinetic under IR-$\varphi_{Auto}$, does not present a depletion of the acceleration because the radiative flux emitted is maximized and continuous.

Under these conditions, the time to gelation and vitrification are reduced to 5 and 25 minutes respectively. From the industrial point of view, this makes it possible to put forward the infrared radiation at low temperatures as a good cure-on-demand way.

### 3.2. Temperature effect

The non-thermal effect is directly related to the radiative flux. Besides, the radiative flux is also indirectly linked to the temperature since the increase in the radiative flux leads to an increase in the temperature of the BADGE-TETA mixture. Consequently, the correlation between the non-thermal effect and the setpoint temperature must be assessed. In order to address this issue, conversions (for maximum radiative flux, *i.e.* IR-$\varphi_{Max}$) are measured versus time at various temperatures: 40 °C, 60 °C and 80 °C. In conjunction with the test at 50 °C already presented, the results of the kinetics are shown in the Figure 4.

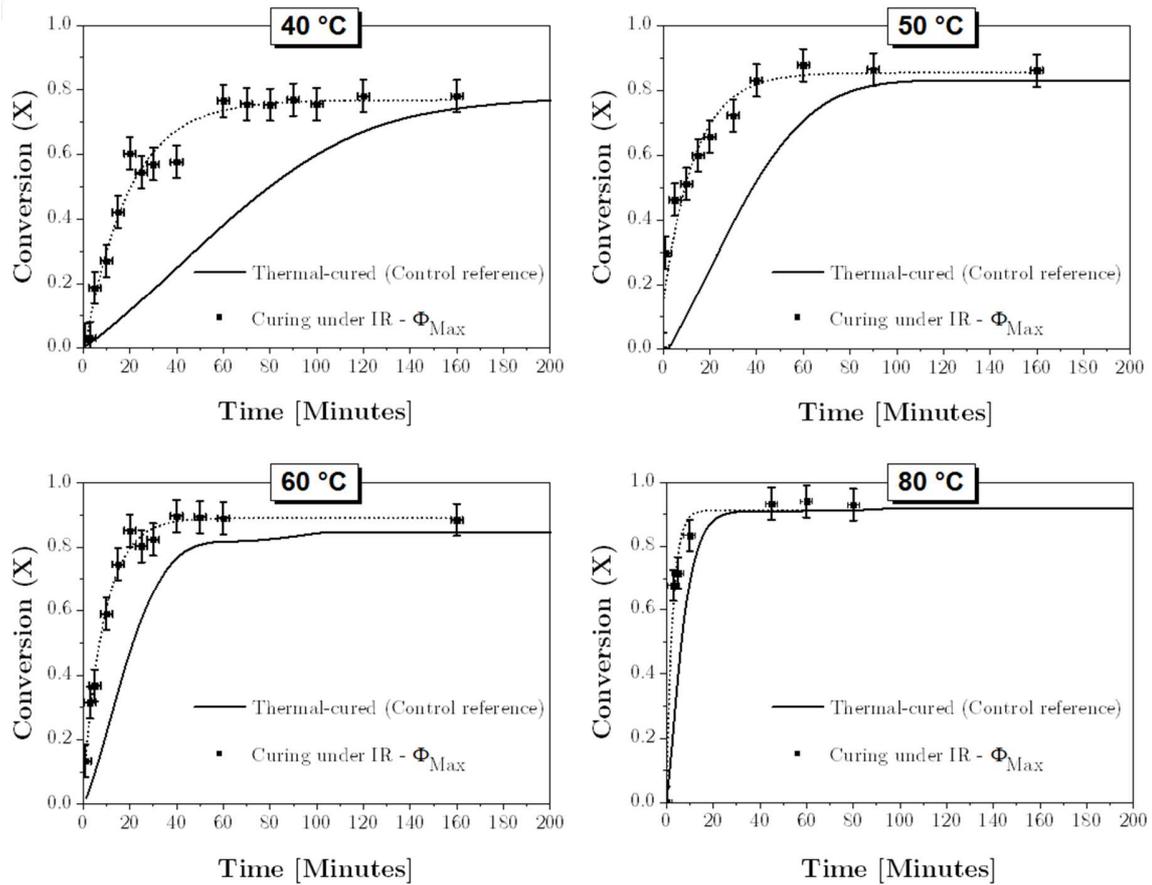

Figure 4 : Conversions of BADGE-TETA mixture cured at various temperature, thermally (-) and with infrared (■); the short-dotted curves ( · · ) are only eye guides.

Basically, in the Figure 4, polymerization kinetics and the maximum conversion increase when the temperature increases. For kinetic with infrared with a maximum radiative flux (IR-$\varphi_{Max}$), the "non-thermal effect" is noticed for all temperatures, but appears to be less and less significant as the temperature increases. Additionally, from 80 °C onwards, this "non-thermal effect" appears to be negligible. Let us note that the same trial at 100 °C was carried out, but that the kinetics of infrared or thermal curing shows exactly the same profile. For greater figure clarity, and because this does not alter the conclusions that follow, these results are not presented. In other words, the thermal and "non-thermal" effects do not seem to be correlated, but rather in competition, leading to the same final network (*i.e.* $T_{g\infty}$ and the infrared spectrum are identical for both thermally and infrared-cured model adhesive). As the temperature increases, the increase in kinetics by thermal effect becomes predominant, completely annihilating the observation of the "non-thermal effect". It appears that this effect has so far never been observed because infrared polymerization has always been studied when the induced temperatures are very high [17–19].

### 3.3. Role of reactive groups during IR-curing

The IR-curing increases the curing kinetics but does not modify the network formed by the reaction between epoxies and amines as the glass transition and the infrared spectra of both networks are the same. The role of temperature in the curing speed is well-known, but the "non-thermal effect" has to be investigated in details and the causes must be identified. One hypothesis that can be formulated is whereby the "non-thermal effect" is triggered by the infrared absorption of reactive groups, leading to a greater reactivity. To support this assumption, a filtration of infrared radiation is proposed (Figure 5). For epoxy-centered infrared curing, a band-pass filter supplied by Edmund Optics (York, United Kingdom) has been used. The filter is a circular Germanium substrate with a thickness of 1 mm and a diameter of 25 mm. It transmits 90 % of the wavelength at 10.6 µm with a half-height width of 1.5 µm (*i.e.* between 830 and 1100 cm⁻¹ with a maximum at 940 cm⁻¹). For reminder, in the mid-infrared range, the stretching C-O of oxirane ring absorb at 915 [34–36] so that this filtration is epoxy-centered. Then, for amine-centered infrared curing, a band-pass filter supplied by Edmund Optics has been used. The filter is a circular sapphire ($Al_2O_3$) substrate with an unknown coating. The filter is 1 mm thick and 25 mm in diameter. It transmits 65 % of the wavelength at 2.95 µm with a half-height width of 0.11 µm (*i.e.* between 3270 and 3520 cm⁻¹ with a maximum at 3390 cm⁻¹). For reminder, in the mid-infrared range, the N-H groups of primary and secondary amines absorb at 3 400 cm⁻¹ respectively [34–36], in such a manner that this filtration is amine-centered.

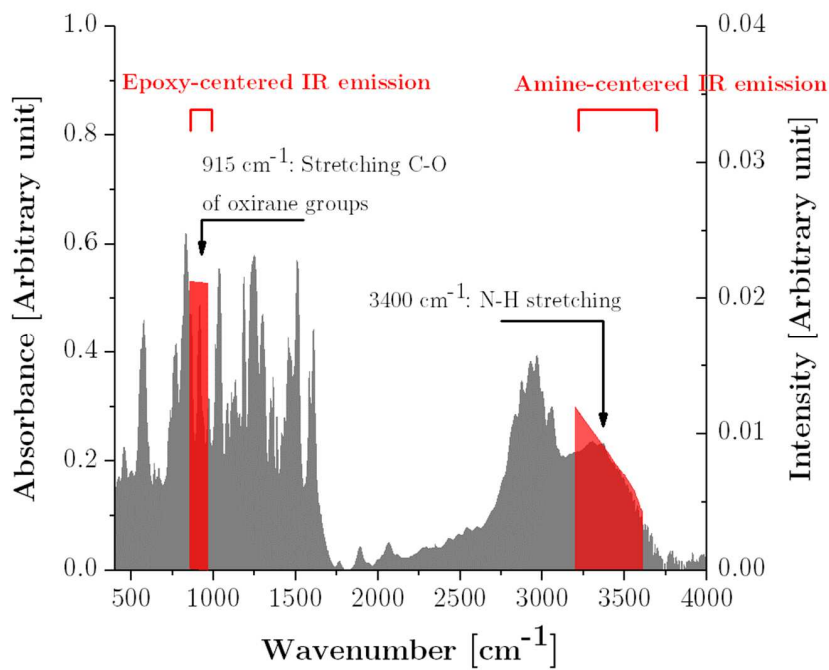

Figure 5: Absorbance of the BADGE-TETA mixture (gray) and emission of the infrared lamp through the epoxy and amine centered filters (red).

For this test, infrared curing was only technically possible if the radiative flux is controlled automatically (IR-$\varphi_{Auto}$). The kinetic follow-up of the model adhesive polymerized with infrared centered around epoxy (★) and amine (■) bands can be compared to the follow-up of unfiltered infrared (○) and pure thermal (-) polymerization (Figure 6).

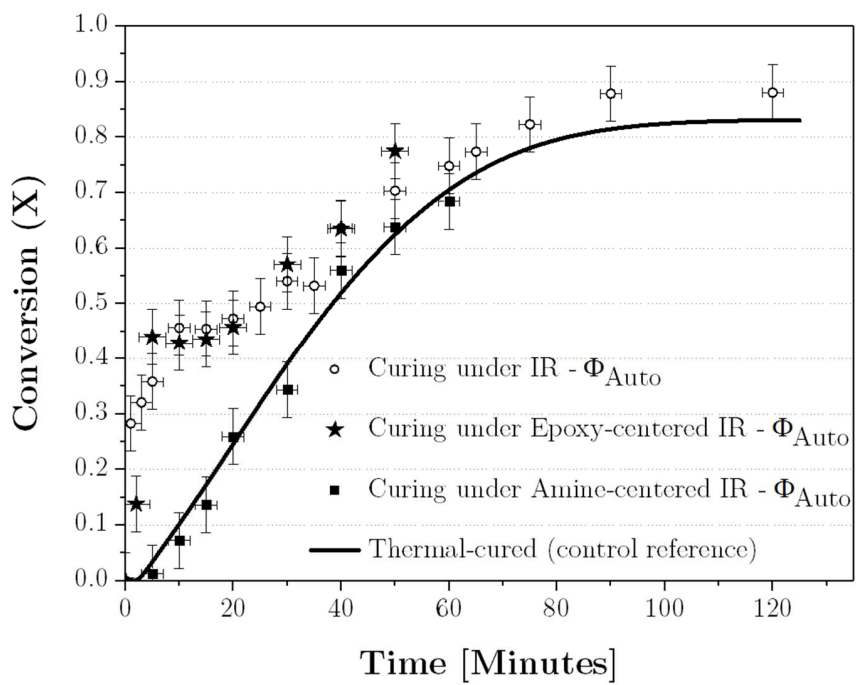

Figure 6 : Conversions of BADGE-TETA mixture cured at 50 °C, thermally (**-**), under IR (○), under epoxy-centered IR (★) and under amine-centered IR (■).

Let us first consider the infrared polymerization kinetics centered around epoxides (★). Compared to unfiltered infrared polymerization, the kinetics is very analogous (taking into account the uncertainty). In a word, whether the infrared spectrum is a broad spectrum (*i.e.* 400-4,500 cm⁻¹) or centered around epoxy groups, the "non-thermal effect" is identical. Then, the kinetic profile of the mixture curing under amine-centered IR is the same than the pure thermal polymerization. Brightly, the "non-thermal" effect of infrared polymerization is no longer noticeable when the radiation is filtered to allow only the infrared radiation that excites the amine groups to pass through.

To put in a nutshell, the "non-thermal" effect admitted so far only exists through the absorption of infrared radiation by epoxy groups. The result is that the epoxy groups, excited by infrared rays, are more reactive. To go further, the non-thermal side of the phenomenon is well proven since for the same amount of energy (conducting at a temperature of 50°C), the kinetics is completely different, depending on the wavelength under consideration. The objective now is to determine how the absorption of infrared radiation by epoxides accelerates the epoxy-amine kinetics.

### 3.4. Causes of the non-thermal effect

The role of epoxy groups has been demonstrated in the "non-thermal effect" of low temperature infrared polymerization. Basically, the epoxy-amine curing occurring in three main reactions is basically accepted (Figure 7).

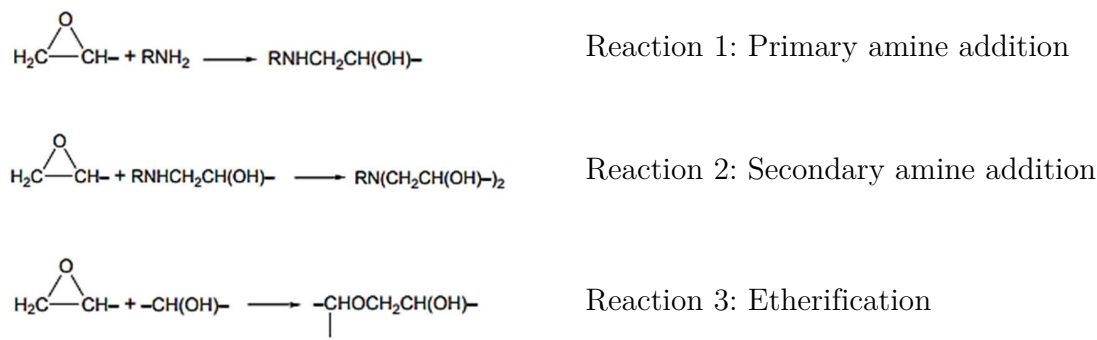

<div>

$H_2C$⎯⎯CH⎯ + $RNH_2$ ⎯⎯→ $RNHCH_2CH(OH)$⎯　　　Reaction 1: Primary amine addition

$H_2C$⎯⎯CH⎯ + $RNHCH_2CH(OH)$⎯ ⎯⎯→ $RN(CH_2CH(OH)⎯)_2$　　　Reaction 2: Secondary amine addition

$H_2C$⎯⎯CH⎯ + ⎯$CH(OH)$⎯ ⎯⎯→ ⎯$CHOCH_2CH(OH)$⎯　　　Reaction 3: Etherification

</div>

Figure 7: Mechanism of non-catalyzed epoxy-amine systems.

In order to separately identify the possible acceleration of each of these three reactions with infrared, a modification of the BADGE-TETA model adhesive is proposed using various hardeners (Figure 8). The tertiary amine is used as an initiator of anionic polymerization of epoxide [30,37–39]. The ethylenediamine has been chosen because it only presents primary amines whereas DMEDA (stereochimically close to EDA) consists only of secondary amines.

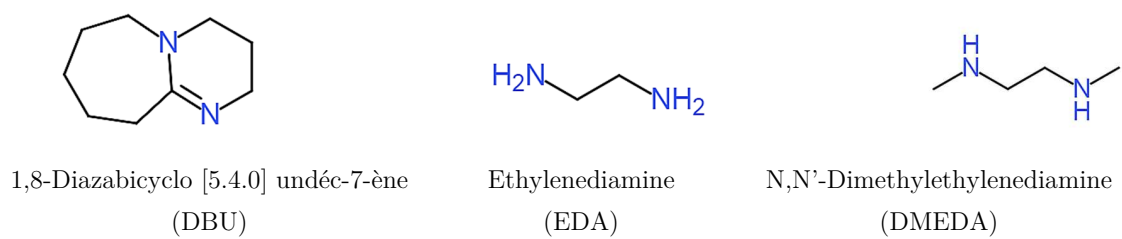

1,8-Diazabicyclo [5.4.0] undéc-7-ène          Ethylenediamine          N,N'-Dimethylethylenediamine
      (DBU)                    (EDA)                (DMEDA)

Figure 8: Chemical structures of DBU, EDA and DMEDA[1].

To confirm the possible acceleration of homopolymerization under infrared, two system modifications are proposed: first, by the epoxy prepolymer alone, then by a BADGE-DBU mixture. DBU is a tertiary amine (extremely strong base - pKa = 11.5), which does not react directly with the epoxy prepolymer but makes the homopolymerization possible [30,37–39]. Tertiary amines are embodied at relatively low proportions (up to 5 phr) as their function are genuinely catalytic [40]. From this bibliographic data and by empirical experience, the chosen amount of DBU is 1.5 phr (*i.e.* 1.5 grams per hundred grams of resin). The homopolymerization kinetics of BADGE alone or catalyzed by DBU are summarized in Figure 9.

---

[1] All the chemical schemes of the molecules were obtained freely on Scribmol.

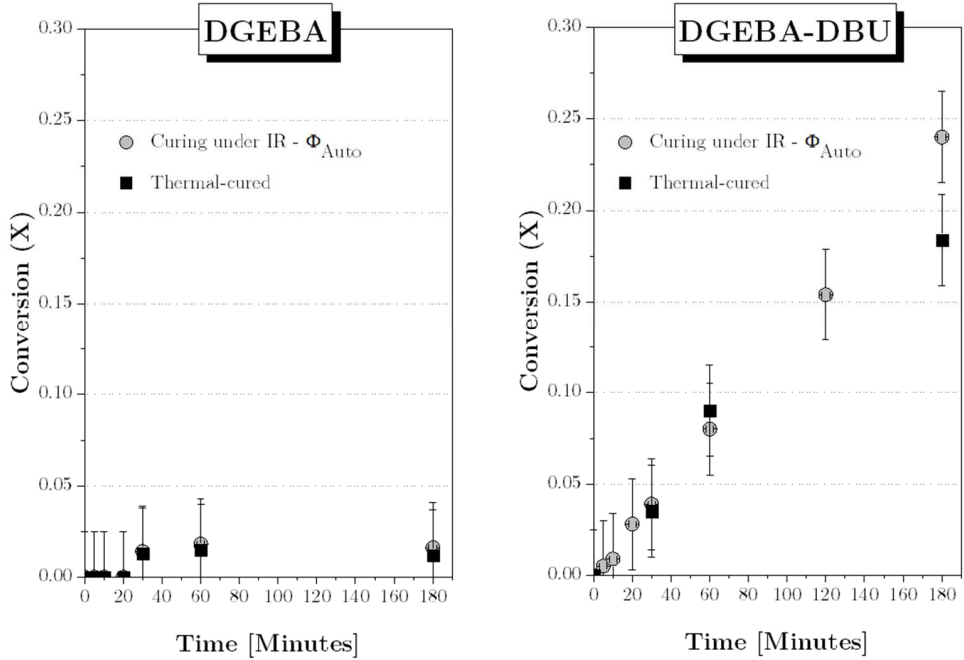

Figure 9 : Conversions of BADGE and BADGE-DBU mixture with infrared.

The homopolymerization of BADGE alone does not occur at 50 °C and is also not observable with infrared (*i.e.* conversion less than 2 % after 3 hour). Regarding the BADGE-DBU mixture, homopolymerization of the epoxy prepolymer occurs from the first hours at 50 °C, but does not seem to be catalyzed with infrared. It is possible that the catalytic effect of infrared radiation may be measurable after 180 minutes, but this is in any case insufficient for the explanation of the thermal effect observed in the first few minutes. In order to verify the acceleration of the epoxy-secondary amine reaction (reaction 2), the kinetic follow-up with infrared of the BADGE polymerized with DMEDA is carried out. the result of the kinetic monitoring with infrared and in pure thermal conditions at 50 °C (Figure 10).

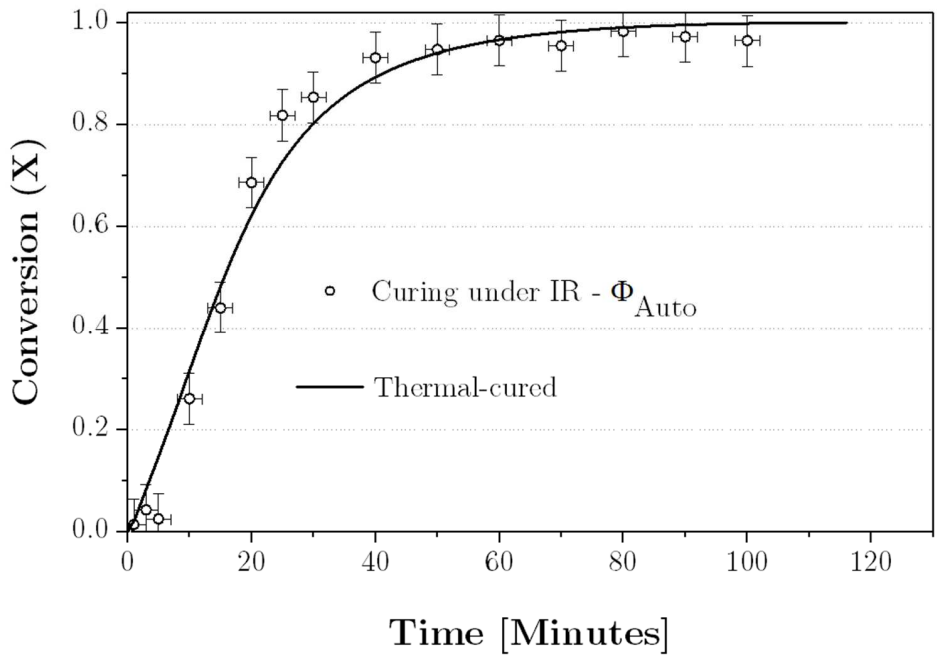

Figure 10 : Polymerization kinetics of the BADGE-DMEDA mixture with infrared.

Both kinetic profiles are, taking account to the uncertainty, the same. In other words, the BADGE-DMEDA reaction has no "non-thermal effect" and is accelerated only by the temperature increase. This means that the secondary epoxy-amine reaction is not catalyzed by the infrared absorption for this system.

The possible acceleration of the primary epoxy/amine reaction (Reaction 1) is studied. For this purpose, the BADGE epoxy prepolymer is polymerized using a hardener with only primary amine: EDA. Obviously, once the primary amines have reacted, secondary amines will be formed and the configuration will be the same as the previous test. But in this case, the proportion of primary amine (in relation to the secondary amines) is much higher so that the acceleration of epoxy-primary amine reaction would be highlighted. The kinetic with infrared is performed and the result (compared to the kinetics of the BADGE-EDA mixture at 50°C) was follow-up (Figure 11).

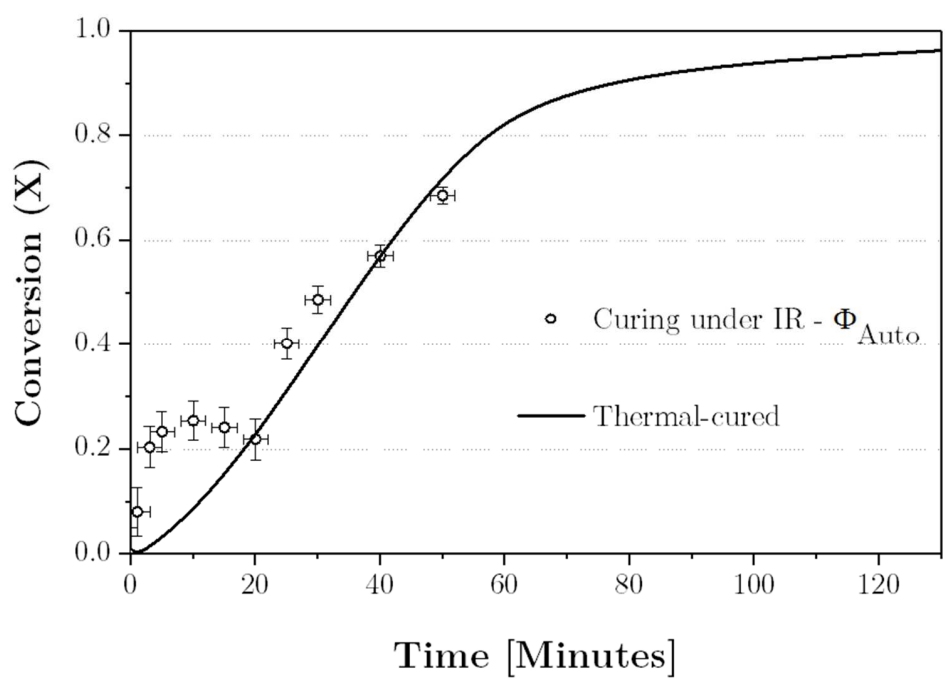

Figure 11 : Polymerization kinetics of the BADGE-EDA mixture with infrared.

The infrared polymerization kinetics of the BADGE-EDA mixture have the same profile as the initial BADGE-TETA mixture tests. Indeed, a strong acceleration of polymerization is observed during the first ten minutes, followed by a slowdown and recovery according to the kinetic profile obtained in thermal. In order to synthesize, the infrared polymerization of different systems has made it possible to highlight that only the epoxy/primary amine reaction shows this "non-thermal effect" with the presented systems. The acceleration of epoxy-epoxy reaction (homopolymerization) or epoxy-secondary amines reaction is only due to an increase in the temperature of the system induced by infrared radiation (thermal effect). Therefore, the "non-thermal effect" of infrared radiation allows a selectivity of the acceleration of the epoxy-amine reaction.

### 3.5. Activation Energy calculation

The purpose of the following is to quantify and propose a more fundamental explanation of the phenomenon. Since conversion monitoring is not enough sufficient, the activation energy calculation of the epoxy-amine reaction is proposed in order to go further. For this first approach, the epoxy-amine reaction is considered as proceeding in a one-step or more precisely, the activation energy considered is that of the epoxy-amine reaction as a whole (both epoxy-primary amine and epoxy-secondary amine).

The polymerization kinetics of epoxy can be presented as a one-step kinetic equation [41,42]:

$$\frac{d\alpha}{dt} = K(T) \times f(\alpha)$$

Equation 6

where $T$ is the temperature, $\alpha$ is the conversion and $f(\alpha)$ is the reaction model. $K(T)$ can take the Arrhenius form so that the Equation 6 can be written as follow:

$$K(T) = k_0 \times e^{-(E/RT)}$$

Equation 7

with $E$ the activation energy. Since the epoxy-primary amine kinetic is catalyzed by infrared irradiation whereas the same reaction with secondary amines is not, a modification of the reaction mechanism can be advanced (*i.e.* with different reaction intermediates). In this case, the activation energy of the epoxy-amine reaction should be different. To calculate this activation energy, the isoconvertional model-free approach was chosen. The model-free approach was adopted as it provides a reliable calculation of the activation energy, even if it varies according to the conversion degree $\alpha$. Simply put, whether the reaction proceeds in one-step ($E$ invariant with $\alpha$) or whether the activation energy is considered as an average value of a multistep reaction, the use of a model free approach will lead to more reliable ad precise result [43]. This model-free approach is based on the assumption that at a given conversion, the polymerization kinetic only depends on the temperature [41,43–45].

$$\left[ \frac{\partial \ln\left( \frac{d\alpha}{dt} \right)}{\partial \; T^{-1}} \right]_\alpha = -\frac{E}{RT} \qquad\qquad \text{Equation 8}$$

Then, for isothermal conditions, which is our study case, the Equation 7 can be modified as:

$$\ln t_{\alpha,i} = -\ln\frac{k_0}{g(\alpha)} + \frac{E}{RT_i} \qquad\qquad \text{Equation 9}$$

where $t_{\alpha,i}$ is the time to reach the considered conversion at a certain temperature $T_i$. From this equation, the slope of the plot $\ln t_{\alpha,i} = f\left( \frac{1}{T} \right)$ gives $\frac{E}{R}$. For the calculation, the temperature considered are 40, 50, 60 and 80 °C. The activation energies can then be calculated as a function of the conversion ($\alpha$) for the thermally polymerized model adhesive, and with infrared.

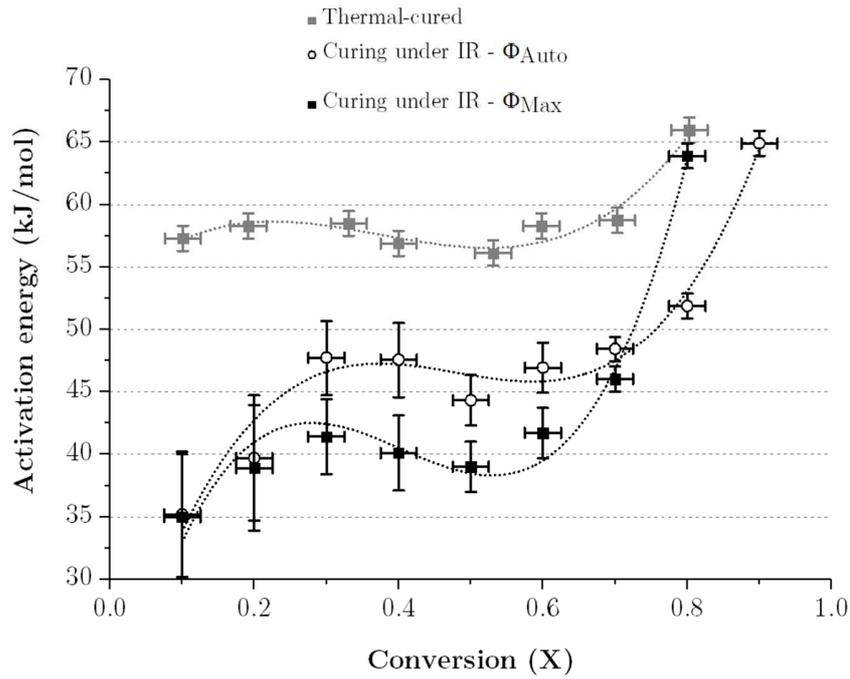

Figure 12: Activation energies of BADGE-TETA cured thermally (■), under IR-$\varphi_{Auto}$ (○) and IR-$\varphi_{Max}$ (■), the short-dotted curves ( · · ) is only an eye guide.

First of all, it should be noted that the activation energy diverges from a conversion of 0.7, and this in the case of the 3 types of polymerization. This divergence is correlated with the fact that the conversions used for the calculation are absolute conversions [12,46]. In fact, for polymerization temperatures below the maximum glass transition temperature (*i.e.* < 138 °C for our model BADGE-TETA system), the maximum conversions are not equal to 1. Therefore, after vitrification of the model adhesive, the calculated activation energy mathematically tends towards infinity. Now let us continue with the activation energies of the thermally cured system (■). For conversions below 0.7, the activation energies are around 57 kJ.mol$^{-1}$ during the polymerization. This value is consistent with the literature on epoxy reactions [45,47,48].

Let us now analyze what is obtained for infrared polymerization. For both IR-$\varphi_{Auto}$ (○) and IR-$\varphi_{Max}$ (■), the activation energies for the low conversions are very low (< 40 kJ.mol$^{-1}$) and increase to reach a plateau of 40 kJ.mol$^{-1}$ for IR-$\varphi_{Max}$ and 47 kJ.mol$^{-1}$ for IR-$\varphi_{Auto}$. In addition, during the entire polymerization process, activation energies remain lower than those obtained in thermal curing. This clearly indicates the catalytic character of the "non-thermal effect".

Unfortunately, one of the limitations of this model is that the starting hypothesis is the one-step epoxy-amine reaction, which does not differentiate between the activation energies of the epoxy-primary amine and epoxy-secondary amine reactions in order to prove that only the epoxy-primary amine reaction is accelerated. On the other hand, this hypothesis can be seen in the fact that the activation energy increases up to a conversion of 0.3, which corresponds perfectly to the decrease in the concentration of the five primary amines.

## 4. CONCLUSION

This paper is focus on infrared curing which, driven by a strong industrial expectation, paves the way for an accelerated polymerization, at low temperature and without initiator addition in the adhesive formulation (*i.e.* universal). The study was carried out on a model mixture of BADGE-TETA for a curing temperature of 50 °C. With infrared, two variations were carried out, one with a radiative flux automatically controlled by the machine (IR-$\varphi_{Auto}$), the other artificially brought to its maximum (IR-$\varphi_{Max}$), and this for an induced temperature, in both cases of 50 °C.

It has been shown that infrared curing is faster than thermal curing at the same temperature. This may have highlighted an effect that, to our knowledge, has never been observed before: simply called hereinbefore "non-thermal effect". This new effect of infrared radiation is not correlated with temperature and seems to be related to the emitted radiative flux. A competition between the thermal and "non-thermal" effects could be observed, so that from a certain temperature, the thermal effect is pre-dominant, completely masking the "non-thermal" effect of infrared irradiation. Therefore, it is only observable at relatively low temperatures, up to 60 °C.

The use of filters to select the lamp's infrared emission around the absorption range of reactive groups has made it possible to show that the "non-thermal" effect of infrared only accelerates the epoxy-primary amine reaction.

By changing the adhesive system, it could be shown that only the epoxy-primary amine reaction was affected by the "non-thermal effect" of infrared irradiation. With an iso-convertional model-free approach, this acceleration has been linked to a reduction in the energy barrier.

Even if this is not the main objective of this paper, low temperature infrared curing would allow, for example, a selectivity of the epoxy-amine reaction which finds applications in many other sectors than in aeronautic.